\documentclass[aps,twocolumn,showpacs,preprintnumbers,amsmath,amssymb]{revtex4}

\usepackage{graphicx}
\usepackage{amsmath}

\renewcommand{\rho}{\varrho}
\renewcommand{\phi}{\varphi}
\renewcommand{\vec}{\mathbf}

\begin{document}

\title{Free-energy functional of the Debye-H\"uckel model of two-component plasmas}

\author{T. Blenski$^1$\footnote{Corresponding author\\Electronic address: thomas.blenski@cea.fr} and R. Piron$^2$}
\affiliation{$^1$Laboratoire ``Interactions, Dynamiques et Lasers'', UMR 9222, CEA-CNRS-Universit\'e Paris-Saclay, Centre d'\'Etudes de Saclay, F-91191 Gif-sur-Yvette Cedex, France.}
\affiliation{$^2$CEA, DAM, DIF, F-91297 Arpajon, France.}

\date{\today}

\begin{abstract}
We present a generalization of the Debye-H\"uckel free-energy-density functional of simple fluids to the case of two-component systems with arbitrary interaction potentials. It allows one to obtain the two-component Debye-H\"uckel integral equations through its minimization with respect to the pair correlation functions, leads to the correct form of the internal energy density, and fulfills the virial theorem. It is based on our previous idea, proposed for the one-component Debye-H\"uckel approach, and which was published recently \cite{Piron16}. We use the Debye-Kirkwood charging method in the same way as in \cite{Piron16}, in order to build an expression of the free-energy density functional. Main properties of the two-component Debye-H\"uckel free energy are presented and discussed, including the virial theorem in the case of long-range interaction potentials.
\end{abstract}


\maketitle

\section{Introduction}
The Debye-H\"uckel (DH) model was introduced  in the theory of electrolytes \cite{DebyeHuckel23}. It is the linearized version of the  Non-Linear Debye-H\"uckel (NLDH) or Poisson-Boltzmann model. Both the DH and NLDH models can be in principle extended to arbitrary interaction potentials, with many applications in the physics of classical fluids, charged or not. 

An important feature of the one-component DH model is the mean-field screening of the interaction potential, that results in the decay of the correlation function, even when the interaction potential has a Coulomb tail. This  property is preserved in the two-component DH theory of plasma or charged mixtures. In general the DH model is valid in the low-coupling limit, i.e. when the interaction energy is small compared to the kinetic energy of particles. The DH theory is a relatively simple mean-field approach to classical fluids in comparison to the Hyper-Netted Chain (HNC) \cite{Morita58} or Percus-Yevick \cite{PercusYevick58} models, which account for part of the correlations. However, several more involved theoretical studies also proceed by introducing corrections, using the low-coupling DH limit as a starting point (see, for instance, \cite{Abe59,DeWitt65}). Recently it was also shown that, in the DH model, the energy and virial "routes" are thermodynamically consistent, for any potential \cite{Santos09}. 
For these reasons, the two-component DH model can be of interest not only for the plasma physics community (see, for instance, \cite{KidderDeWitt61,Vieillefosse81}) but also in the physics of electrolytes (see, for example  \cite{Nordholm84, Penfold90}).
In the case of a long-range attractive potential, the linearization performed in the DH model allows one to circumvent the ``classical catastrophe'' of collapsing particles. To some extent, this explains why this model is so commonly used in plasma physics as well as in the physics of electrolytes.

Our interest in the DH models stems from our research on density effects in the equation of state and in radiative properties of dense plasmas. We aim to describe these effects using approximate but fully variational models of atoms in plasmas. The ion and free electron correlations and their impact on the atomic structure and dynamics should thus be taken into account while preserving the variational character of the approach. Up to now the most used models are still essentially based on the ion-in-cell picture \cite{Feynman49,Rozsnyai72,Liberman79}. Some progress towards a variational formulation of quantum atom-in-plasma models was achieved \cite{Blenski07b, Blenski07a, Piron11, Piron11b, Blenski13, Piron13, Caizergues14, Caizergues16}. However, it relies on a very simple hypothesis on ion-ion correlations. Attempts to include ion correlations into atom-in-plasma models were also proposed (see, for instance, \cite{Ofer88, Starrett12, Starrett12b, Chihara16}), but do not stem from a fully variational derivation.

The DH model is a natural candidate tool in order to include plasma ions and electron effects for relatively weakly coupled plasma. The results presented in this paper extend the variational free-energy density formula presented in \cite{Piron16} to the two-component case. The present DH theory can also be useful to construct variational approaches to models of two-component fluids using expressions of the free-energy density as a functional of the pair correlation functions. In the atom-in-plasma models such a variational expression can be used in a more general theory that also includes the ion electronic structure. 

A two-component free-energy density functional is available in the HNC theory \cite{Lado73b, Enciso87}. However, as it was the case in the well-known one-component DH theory, an expression of the free-energy density as a functional of the pair correlation functions has not been yet proposed. The purpose of this paper is to present a brief derivation of such an expression in the DH approximation, which can be seen as the DH equivalent to the HNC free-energy density functional of Lado \cite{Lado73b} (see also \cite{Enciso87}).

\section{Debye-H\"uckel integral equations of a two-component fluid}

Let us consider a two-component homogeneous fluid at equilibrium, at a temperature $T=1/(k_B\beta) $. In the case of plasmas or charged liquids the fluid neutrality results in the charge balance $\sum_{j}\rho_j\:z_j=0$ where $ \rho_j $ is the average particle density and $ z_j $ the charge of the $j$ specie, respectively.
The particles interact through potentials  $u_{ij}(r)$. Using the so called ``Percus trick'' (see \cite{Percus64}, also \cite{HansenMcDonald}) we obtain the following identity :

\begin{equation}\label{eq_Percus} 
\frac{\rho^{(1:j)}\left\{\{\phi_n(r')=u_{in}(r')\};r\right\}}{\rho_j}=g_{ij}(r)
\end{equation}
where the $g_{ij}(r)$ are the pair distribution functions of the homogeneous fluid, for the species $i$ and $j$, and $ \rho^{(1:j)}(r)$ is the $j$-specie 1-body reduced density matrix for a non-homogeneous fluid, with an external potential $\phi_n(r')$ acting on each specie $n$.

The DH model is obtained from the static linear response of the density to the external potential:
\begin{align}
&\rho^{(1:j)}\left\{u_{in}(r'');r\right\}\approx\rho_j \nonumber
\\
&+\sum_{n}\int d^3r'\left\{
\left.
\frac{\delta \rho^{(1:j)}\left\{\{\phi_n(r'')\};r\right\}}{\delta \phi_n(r')
}\right|_{\{\phi_n(r')=0\}}
u_{in}(r')
\right\}
\end{align}\label{eq_LR}
The functional derivatives of the density are  (Yvon equations):
\begin{align}
\frac{1}{\beta}
\frac{\delta \rho^{(1:j)}(\vec{r})}{\delta \phi_j(\vec{r}')}
=&
-\rho^{(2:j)}\left(\vec{r},\vec{r}'\right)
+\rho^{(1:j)}\left(\vec{r}\right)
\rho^{(1:j)}\left(\vec{r}'\right)
\nonumber\\
&-\rho^{(1:j)}\left(\vec{r}'\right)
\delta_3(\vec{r}-\vec{r}')\label{eq_derivative_fi_j}
\\
\frac{1}{\beta}
\frac{\delta \rho^{(1:j)}(\vec{r})}{\delta \phi_{n\neq j}(\vec{r}')}
=&
-\rho^{(1:j,1:n)}\left(\vec{r},\vec{r}'\right)
+\rho^{(1:j)}\left(\vec{r}\right)
\rho^{(1:n)}\left(\vec{r}'\right)\label{eq_derivative_fi_n}
\end{align}
From the above equations and the definition (see, for instance, \cite{HansenMcDonald}) of the correlation functions $h_{ij}(r)+1=g_{ij}(r) $ one obtains the equations of the Debye-H\"uckel model :
\begin{equation}\label{eq_DH_ij}
h_{ij}(r)=
-\beta u_{ij}(r)-\beta\sum_{n} \rho_n\int d^3r'\left\{
u_{in}(r')h_{in}(|\vec{r}-\vec{r}'|)
\right\}
\end{equation}
For the purpose of future considerations it is useful to write the DH equations in the following symmetrical form:
\begin{align}
h_{11}(r)=
-\beta u_{11}(r)
-\beta\int &d^3r'\left\{
\rho_1u_{11}(r')h_{11}(|\vec{r}-\vec{r}'|)
\right.\nonumber\\ 
&\left.
-\rho_2u_{12}(r')h_{12}(|\vec{r}-\vec{r}'|)
\right\}\label{eq_DH_11}
\\
h_{22}(r)=
-\beta u_{22}(r)
-\beta\int &d^3r'\left\{
\rho_2u_{22}(r')h_{22}(|\vec{r}-\vec{r}'|)
\right.\nonumber\\ 
&\left.
-\rho_1u_{12}(r')h_{12}(|\vec{r}-\vec{r}'|)
\right\}\label{eq_DH_22}
\end{align}
\begin{align}
h_{12}(r)=&
-\beta u_{12}(r)
\nonumber\\
-\frac{\beta}{2}
\int &d^3r'\left\{
u_{12}(|\vec{r}-\vec{r}'|)
\left(\rho_1h_{11}(r')
+\rho_2h_{22}(r')\right)
\right.\nonumber\\
&\left.
+h_{12}(|\vec{r}-\vec{r}'|)
\left(\rho_1u_{11}(r')
+\rho_2u_{22}(r')\right)
\right\}\label{eq_DH_12}
\end{align}

We define the Fourier transform $\mathcal{F}_{\vec{k}}$ of a function $\mathcal{F}(\vec{r})$ as $\mathcal{F}_{\vec{k}}=\int d^3r\left\{\mathcal{F}(\vec{r})e^{i\vec{k}.\vec{r}}\right\} $. The DH equations have a simple form in the Fourier space:
\begin{align}
h_{11;k} + \rho_1 h_{11;k}\, \beta u_{11;k} + \rho_2 h_{12;k}\, \beta u_{12;k}&=-\beta u_{11;k}
\label{DH_2_k_11}
\\
h_{22;k} + \rho_1 h_{12;k}\,\beta u_{12;k} + \rho_2 h_{22;k}\,\beta u_{22;k}&=-\beta u_{22;k}
\label{DH_2_k_22} 
\\
h_{12;k} + \rho_1 h_{11;k}\,\beta u_{12;k} + \rho_2 h_{12;k}\,\beta u_{22;k}&=-\beta u_{12;k}
\label{DH_2_k_12}
\end{align} 
The solutions of these equations are:
\begin{align} 
h_{\text{eq},11;k}&=-\frac{1}{\rho_1}+\frac{1}{\rho_1 \beta}\frac{\beta+{\beta}^{2}\rho_2 u_{22;k}}{D_k}
\label{F_E_h11}
\\
h_{\text{eq},22;k}&=-\frac{1}{\rho_2}+\frac{1}{\rho_2 \beta}\frac{\beta+{\beta}^{2}\rho_1 u_{11;k}}{D_k}
\label{F_E_h22}
\\
h_{\text{eq},12;k}&=-\frac{1}{\rho_1\rho_2\beta}\frac{{\beta}^{2}\rho_1\rho_2 u_{12;k}}{D_k}
\label{F_E_h12}
\end{align}
where we have defined:
\begin{equation}\label{def_D} 
D_k=1+\beta(\rho_1 u_{11;k}+\rho_2 u_{22;k}) +{\beta}^{2} \rho_1 \rho_2 (u_{11;k}u_{22;k}-u_{12;k}^{2})
\end{equation}

\section{Expression for a free-energy functional of a two-component fluid}

We are looking for a functional of trial correlations functions $h_{ij}(r)$ which, when minimized with respect to these functions,  gives the DH equations and, at the DH equilibrium, has the value of the free-energy excess due to the interactions. As in \cite{Lado73,Lado73b,Piron16}, we use the charging procedure due to Debye and Kirkwood \cite{Kirkwood35}. The charging parameter $\xi$ allows one to switch on the interaction potentials from zero to their actual values $u^{\xi}_{ij}(r)=\xi\,u_{ij}(r)$.
For an exact interacting system one gets from the grand canonical statistical sum (see, for instance, \cite{HansenMcDonald}) the following exact expression for the free-energy excess per unit volume:
\begin{align}\label{eq_charging_A_2}
&\frac{A_\text{eq}^\xi\left\{{\{\rho}_{i}\},\beta,\{{u}_{ij}(r)\}\}\right\}}{V}
\nonumber\\
&=\int_0^\xi\frac{d\xi'}{\xi'}\int d^3r\left\{\frac{1}{2}\sum_{i,j}\rho_i \rho_j h_{\text{eq},ij}^{\xi'}(r)u_{ij}^{\xi'}(r)\right\}
\end{align}

We require, as in \cite{Lado73,Lado73b,Piron16}, that in our approximate two-component DH theory, the searched free-energy density functional gives, at the equilibrium, the value of Eq.~\eqref{eq_charging_A_2}, with DH equilibrium functions $ h_{\text{eq},ij}(r)$ (or their equivalent forms in the Fourier space). Moreover, as in \cite{Piron16}, we postulate that the searched functional, which depends on arbitrary trial functions $h_{ij;k}$, can be written as follows in the Fourier space:
\begin{widetext} 
\begin{align}
\label{eq_A_form_postulated}
\frac{A\left\{\{\rho_i\},\beta,\{u_{ij}(r)\},\{h_{ij}(r)\}\right\}}{V}=
\int\frac{d^3k}{(2\pi)^3}\left\{
\frac{f_k}{2} 
\left[
\rho_1^2
\left(\frac{h_{11;k}^2}{2}
+ \beta u_{11;k} h_{11;k}
+ \rho_1\beta u_{11;k} \frac{h_{11;k}^2 }{2}
+\rho_2\beta u_{12;k} h_{11;k}h_{12;k}
\right)
\right.\right.\nonumber\\
+
\rho_2^2
\left(\frac{h_{22;k}^2}{2}
+ \beta u_{22;k} h_{22;k}
+ \rho_2\beta u_{22;k} \frac{h_{22;k}^2 }{2}
+\rho_1\beta u_{12;k} h_{22;k}h_{12;k}
\right)
\nonumber\\
\left.\left.
+
2\rho_1\rho_2
\left(\frac{h_{12;k}^2}{2}
+ \beta u_{12;k} h_{12;k}
+ \frac{h_{12;k}^2 }{2}
\frac{1}{2}\left(\rho_1\beta u_{11;k}+\rho_2\beta u_{22;k}\right)
\right)
\right]
\right\}
\end{align}
\end{widetext} 
where the function $f_k$ depends on all $\rho_i$,  $\{u_{ij;k}\}$ and on $\beta$, but not on the $\{h_{ij;k}\}$.
It is easy to check that the functional derivatives of $ A\left\{\{\rho_i\},\beta,\{u_{ij}(r)\},\{h_{ij}(r)\}\right\}$ with respect to the $\{h_{ij;k}\}$, at fixed $\rho_i$ and $\{u_{ij;k}\}$, lead to the DH Eqs.~\eqref{DH_2_k_11}, \eqref{DH_2_k_22} and \eqref{DH_2_k_12}.

We postulate that such a function $f_k$ exists and that the functional of Eq.~\eqref{eq_A_form_postulated}, taken at equilibrium, has the value given in Eq~\eqref{eq_charging_A_2}. A relatively simple way to obtain $f_k$, proving also its existence, is to use the solution of the DH equations in the Fourier space: $h_{\text{eq},ij;k}^\xi$.
Using Eqs.~\eqref{F_E_h11},\eqref{F_E_h22},\eqref{F_E_h12} and \eqref{def_D}, we can recast the free energy-density excess of Eq.~\eqref{eq_charging_A_2} as follows:
\begin{align}
&\frac{A_\text{eq}^\xi\left\{{\{\rho}_{i}\},\beta,\{{u}_{ij}\}\}\right\}}{V}
\nonumber\\
&=\int_0^\xi\frac{d\xi'}{\xi'}\int\frac{d^3k}{(2\pi)^3}\left\{\frac{1}{2}
\sum_{i,j}\rho_i\,\rho_ih_{\text{eq},ij;k}^{\xi'}u_{ij;k}^{\xi'}\right\}\label{eq_charging_A_2_expl_1}
\\
&=\int_0^\xi\frac{d\xi'}{\xi'}\int\frac{d^3k}{(2\pi)^3}\left\{\frac{1}{2\beta}\left[-\beta\left(\rho_1u^{\xi'}_{11;k}+\rho_2u^{\xi'}_{22;k}\right)
\vphantom{\frac{\left(D^{\xi'}_{k}\right)}{D^{\xi'}_{k}}}
+\right.\right.\nonumber\\&\left.\left.
\frac{\beta\left(\rho_1u^{\xi'}_{11;k}+\rho_2u^{\xi'}_{22;k}\right)+2\beta\rho_1\rho_2\left(u^{\xi'}_{11;k}u^{\xi'}_{22;k}-u^{\xi'\,2}_{22;k}\right)}{D^{\xi'}_{k}}\right]\right\}\label{eq_charging_A_2_expl_2}
\\
&=\int_0^\xi\frac{d\xi'}{\xi'}\int\frac{d^3k}{(2\pi)^3}\left\{\frac{1}{2\beta}\left[-\xi'\frac{\partial}{\partial\xi'}\left[\beta\left(\rho_1u^{\xi'}_{11;k}+\rho_2u^{\xi'}_{22;k}\right)
\right.\right.\right.\nonumber\\&
\hphantom{=\int_0^\xi\frac{d\xi'}{\xi'}\int\frac{d^3k}{(2\pi)^3}}
\left.\left.\left.
-\ln\left(D^{\xi'}_{k}\right)\right]\right]\right\}\label{eq_charging_A_2_expl_3}
\end{align}
Changing the order of the integrations we immediately get:
\begin{align}
&\frac{A_\text{eq}^\xi\left\{{\{\rho}_{i}\},\beta,\{{u}_{ij}\}\}\right\}}{V}
\nonumber\\
&=\int\frac{d^3k}{(2\pi)^3}\left\{\frac{1}{2\beta}\left[\ln\left(D^{\xi}_{k}\right)-\left[\beta\left(\rho_1u^{\xi}_{11;k}+\rho_2u^{\xi}_{22;k}\right)\right]\right]\right\}\label{eq_charging_A_2_expl_4}
\end{align}
On the other hand, using the DH equations~\eqref{DH_2_k_11},\eqref{DH_2_k_22} and \eqref{DH_2_k_12}
we can also calculate the  value of the excess free-energy density at the equilibrium from Eq.~\eqref{eq_A_form_postulated}. We obtain:
\begin{align}
&\frac{A_\text{eq}^\xi\left\{{\{\rho}_{i}\},\beta,\{{u}_{ij}\}\}\right\}}{V}
\nonumber\\
&=\int\frac{d^3k}{(2\pi)^3}\left\{\frac{f_{k}}{2}\left[\frac{{\rho_1}^{2}}{2}  h^{\xi}_{\text{eq},11;k}\beta u^{\xi}_{11;k}
+\rho_1\rho_2 h^{\xi}_{\text{eq},12;k}\beta u^{\xi}_{12;k}
\right.\right.\nonumber\\&\left.\left.
\hphantom{=\int\frac{d^3k}{(2\pi)^3}}
+\frac{{\rho_2}^{2}}{2}  h^{\xi}_{\text{eq},22;k}\beta u^{\xi}_{22;k}\right]\right\}\label{eq_A_postulated_eq_1}
\end{align}
Substituting the values of the equilibrium correlation functions given in Eqs.~\eqref{DH_2_k_11},
\eqref{DH_2_k_22}, \eqref{DH_2_k_12} and \eqref{def_D}, we get: 
\begin{align}
&\frac{A_\text{eq}^\xi\left\{{\{\rho}_{i}\},\beta,\{{u}_{ij}\}\}\right\}}{V}
=\int\frac{d^3k}{(2\pi)^3}\left\{\frac{f_{k}}{2}\left[\left(1-\frac{1}{D^{\xi}_{k}}\right)
\right.\right.\nonumber\\&\left.\left.
-\frac{\beta}{2}\left(\rho_1u^{\xi}_{11;k}+\rho_2u^{\xi}_{22;k}\right)\left(1+\frac{1}{D^{\xi}_{k}}\right)\right]\right\}\label{eq_A_postulated_eq_2}
\end{align}
Comparing Eqs.~\eqref{eq_A_postulated_eq_2} and \eqref{eq_charging_A_2_expl_4} we get the searched expression for the function $f_k$:
\begin{equation}
f_k=\frac{\ln\left(D^{\xi}_{k}\right)-\left[\beta\left(\rho_1u^{\xi}_{11;k}+\rho_2u^{\xi}_{22;k}\right)\right]}{\beta\left[\left(1-\frac{1}{D^{\xi}_{k}}\right)-\frac{\beta}{2}\left(\rho_1u^{\xi}_{11;k}+\rho_2u^{\xi}_{22;k}\right)\left(1+\frac{1}{D^{\xi}_{k}}\right)\right]}\label{eq_f_k}
\end{equation}
The expression for the free-energy density functional Eq.~\eqref{eq_A_form_postulated} with the calculated expression for the function $f_k$, Eq.~\eqref{eq_f_k}, is the main result of the present paper.

It is interesting to check whether we can recover from the obtained free-energy density functional Eq.~\eqref{eq_A_form_postulated} the expression for the free-energy density functional of the one-component DH system of \cite{Piron16}. Indeed, if we suppress the interaction between the two different species: $u_{12;k} = 0$ (which also implies $h_{\text{eq},12;k} = 0$), and use  Eq.~\eqref{eq_A_form_postulated} with Eq.~\eqref{eq_f_k}, we get: 
\begin{align}\label{functional_u12zero}
&\frac{A^\xi\left\{{\{\rho}_{i}\},\beta,\{u_{ii}\},\{h_{ii}\}\}\right\}}{V}
\nonumber\\
&= 
\int\frac{d^3k}{(2\pi)^3}\left\{\frac{1}{{\beta}^2}\frac{
\sum_i\left(\ln\left(1+\beta\rho_i u^{\xi}_{ii;k}\right)-\beta\rho_i u^{\xi}_{ii;k}\right)}
{\sum_i\rho_{i}^{2}h_{\text{eq},ii;k}u_{ii;k}} 
 \right.
\nonumber\\
&\times\left.\left[\sum_i\rho_{i}^2\left(\frac{h^{2}_{ii;k}}{2}+h_{ii;k} \beta u_{ii;k}+\rho_i\frac{h_{ii;k}}{2}\beta u_{ii;k}\right)\right]
\vphantom{\frac{\left(u^{\xi}_{ii;k}\right)}{\rho_i^2}}
\right\}
\end{align}
where 
\begin{equation}\label{one_comp_h_eq}
h_{\text{eq},ii;k}=\frac{-\beta u_{ii;k}}{1+\rho_i \beta u_{ii;k}}
\end{equation}
The obtained functional Eq.~\eqref{functional_u12zero}, when minimized with respect to the trial functions $h_{ii;k}$, gives the one-component DH equation for the system $i$. If we substitute into Eq.~\eqref{functional_u12zero} the equilibrium values of Eq.~\eqref{one_comp_h_eq}, then the functional takes the value of the sum of the free energies of the two non-interacting systems $1$ and $2$.

The postulated formula Eq.~\eqref{eq_charging_A_2} is equivalent to the following formulas for the functional derivatives of the excess free-energy density functional with respect to the interaction potentials:
\begin{align}
\left.{\frac{\delta}{\delta u_{11}(r)}} 
\frac{ A\left\{\{\rho_i\},\beta,\{u_{ij}(r)\},\{h_{ij}(r)\}\right\} }{V}
\right|_{\text{eq}}
&=\frac{\rho_1^2}{2}h_{\text{eq},11}(r)\label{det_u11} 
\\
\left.{\frac{\delta}{\delta u_{22}(r)}} 
\frac{ A\left\{\{\rho_i\},\beta,\{u_{ij}(r)\},\{h_{ij}(r)\}\right\} }{V}
\right|_{\text{eq}}
&=\frac{\rho_2^2}{2}h_{\text{eq},22}(r)\label{det_u22} 
\\
\left.{\frac{\delta}{\delta u_{12}(r)}} 
\frac{ A\left\{\{\rho_i\},\beta,\{u_{ij}(r)\},\{h_{ij}(r)\}\right\} }{V}
\right|_{\text{eq}}
&=\rho_1\rho_2h_{\text{eq},12}(r)\label{det_u12} 
\end{align}
where the $|_\text{eq}$ denote values taken for $\{h_{ij}(r)=h_{\text{eq},ij}(r)\}$.
Eqs.~\eqref{det_u11},\eqref{det_u22},\eqref{det_u12} are exact for an exact two-component system (see, for instance, \cite{Evans79}). In our approximate DH model they result from Eq.~\eqref{eq_charging_A_2}. In order to prove it, we rewrite Eq.~\eqref{eq_charging_A_2} in a slightly different way, obtaining: 
\begin{align}
&\frac{A_\text{eq}^\xi\left\{{\{\rho}_{i}\},\beta,\{{u}_{ij}(r)\}\right\}}{V}
\nonumber\\ 
&=\int_0^\xi\frac{d\xi'}{\xi'}\int d^3r
\left\{\frac{1}{2}
\sum_{i,j}\rho_i\rho_j h_{\text{eq},ij}^{\xi'}(r)u_{ij}^{\xi'}(r)\right\}
\nonumber \\
&=\int_0^\xi d\xi'\int d^3r
\left\{\frac{1}{2}
\sum_{i,j}\rho_i\rho_j h_{\text{eq},ij}^{\xi'}(r)\frac{\partial u_{ij}^{\xi'}(r)}{\partial \xi'}\right\}\label{eq_charging_rew}
\end{align}
Let us consider the functional $A^\xi\{{\{\rho}_{i}\},\beta,\{{u}^{\xi}_{ij}(r)\},\{{h}_{ij}(r)\}\}/V$
as given by Eq.~\eqref{eq_A_form_postulated}, where we put $u^{\xi}_{ij}(r)$ instead of $u_{ij}(r)$. As stems from Eq.~\eqref{eq_f_k}, the function $f_k$ depends on $\xi$ only via the potentials $u^{\xi}_{ij}(r)$. Moreover, the functional derivatives of $A^\xi\{{\{\rho}_{i}\},\beta,\{{u}^{\xi}_{ij}(r)\},\{{h}_{ij}(r)\}\}/V$ with respect to $h_{ij}(r)$ are equal to zero at the DH equilibrium. One can thus write formally:
\begin{align}\label{identity_charging}
&\frac{A_\text{eq}^\xi\left\{{\{\rho}_{i}\},\beta,\{{u}_{ij}(r)\}\right\}}{V} 
\nonumber\\
&=\left.\frac{A\{{\{\rho}_{i}\},\beta,\{{u}_{ij}^\xi(r)\},\{{h}_{ij}(r)\}\}}{V}
\right|_{\text{eq},\xi} 
\nonumber\\
&= \int_0^\xi d\xi'\left\{\frac{d}{d\xi'}\left.\frac{A\{{\{\rho}_{i}\},\beta,\{{u}_{ij}^{\xi'}(r)\},\{{h}_{ij}(r)\}\}}{V}\right|_{\text{eq},\xi'}\right\} 
\nonumber\\
&=\int_0^\xi d\xi'\int d^3r
\nonumber\\
&\left\{\frac{\partial}{\partial u_{11} }\left.\frac{A\{{\{\rho}_{i}\},\beta,\{{u}_{ij}^{\xi'}(r)\},\{{h}_{ij}(r)\}\}}{V}\right|_{\text{eq},\xi'}
\frac{\partial u_{11}^{\xi'}(r)}{\partial\xi'} \right. 
\nonumber \\
&\left.
+\frac{\partial}{\partial u_{12} }\left.\frac{A\{{\{\rho}_{i}\},\beta,\{{u}_{ij}^{\xi'}(r)\},\{{h}_{ij}(r)\}\}}{V}\right|_{\text{eq},\xi'}
\frac{\partial u_{12}^{\xi'}(r)}{\partial\xi'} \right.
\nonumber \\
&\left.
+\frac{\partial}{\partial u_{22} }\left.\frac{A\{{\{\rho}_{i}\},\beta,\{{u}_{ij}^{\xi'}(r)\},\{{h}_{ij}(r)\}\}}{V}\right|_{\text{eq},\xi'}
\frac{\partial u_{22}^{\xi'}(r)}{\partial\xi'} \right\}
\end{align}
where the $|_{\text{eq},\xi}$ denote values taken for $\{h_{ij}(r)=h^{\xi}_{\text{eq},ij}(r)\}$.
Comparing Eqs.~\eqref{eq_charging_rew} and \eqref{identity_charging} and taking into account that they are valid for a large class of independent potentials $u^{\xi}_{ij}(r)$, we get Eqs.~\eqref{det_u11}, \eqref{det_u22}, \eqref{det_u12}. These equations can be also verified directly using the explicit form of Eqs.~\eqref{eq_A_form_postulated} and Eq.~\eqref{eq_f_k} but the necessary algebra is tedious.

\section{Internal energy and virial theorem in the case of  two-component DH theory} 

The internal-energy density, as defined in thermodynamics is:
\begin{equation}
\frac{U^\xi_\text{eq}}{V}=\frac{\partial}{\partial \beta}\left(\frac{\beta A^\xi_\text{eq}}{V}\right)=\left.\frac{\partial}{\partial \beta}\left(\frac{\beta A^\xi}{V}\right)\right|_\text{eq}
\end{equation}
As stems from Eqs.~\eqref{eq_A_form_postulated} and \eqref{eq_f_k}, $\beta A^\xi/V$ is a functional of $\{h_{ij}(r)\}$, $\{\rho_i\}$, and of the functions $\{\beta \xi u_{ij}(r)\}$. As a consequence, we can write:
\begin{equation}
\frac{\partial}{\partial \beta}\left(\frac{\beta A^\xi}{V}\right)=\xi\frac{\partial}{\partial \xi}\left(\frac{A^\xi}{V}\right)
\end{equation}
We then have, in virtue of Eq.~\eqref{eq_charging_A_2}:
\begin{align}
&\frac{ U_\text{eq}^\xi \{ \{\rho_{i}\}, \beta, \{u_{ij}(r)\} \} }{V}
\nonumber\\
&=\int d^3r\left\{\frac{1}{2}\sum_{i,j}\rho_i\rho_j h_{\text{eq},ij}^{\xi}(r)u_{ij}^{\xi}(r)\right\}
\label{internal_energy_2}
\end{align}
which corresponds to the exact expression of the internal-energy density, with $h_{\text{eq},ij}^{\xi}(r)$ taken in the DH approximation to the equilibrium correlation functions.

The virial theorem for the exact two-component classical fluid at equilibrium can be derived from the statistical sum in the grand canonical ensemble formalism. We consider this ensemble in a finite volume and the virial pressure is calculated from the exact thermodynamic formula, following the technique used in \cite{hill1956}, Sec.~30. The obtained  expression for the total thermodynamic pressure is $P = P_{0}+P_\text{v}$, where $P_{0}=\left(\rho_1+\rho_2\right)/\beta$ corresponds to the ideal gas formula, while:
\begin{align}
&P_\text{v}\left\{{\{\rho}_{i}\},\beta,\{{u}_{ij}\}\}\right\}
\nonumber\\
&=-\frac{1}{6}\int d^3r\left\{
\sum_{i,j}\rho_i\rho_j g_{\text{eq},ij}(r)\,r\frac{\partial}{\partial r} \left(u_{ij}(r)\right)\right\}\label{Pressure_virial}
\end{align}
is the Clausius virial pressure. We recall that $g_{\text{eq},ij}(r)=1+h_{\text{eq},ij}(r)$. Since we have:
\begin{equation} \label{rV} 
\int d^3r \left\{ r\frac{d}{dr}u_{ij}(r)\right\}
=\lim_{k\rightarrow 0}\left(-3 u_{ij;k}-k\frac{\partial u_{ij;k}}{\partial k}\right)
\end{equation}
we get in the case of interaction potentials having Coulomb tails:
\begin{align}
&\int d^3r\left\{\sum_{i,j}\frac{{\rho_i \rho_j}}{2}\,r\frac{\partial u_{ij}(r)}{\partial r} \right\}
\nonumber\\
&=-\lim_{k\rightarrow 0}\sum_{i,j}\frac{4\pi\rho_i \rho_j z_{i} z_{j}}{2k^{2}}
= -\lim_{k\rightarrow 0}\frac{2\pi}{k^{2}}\left (\sum_j\rho_j z_j\right )^{2}
=0\label{Aux_1}
\end{align}
due to the overall neutrality $\sum_j\rho_j\:z_j=0$. Thus in the case of potentials having Coulomb tails (plasmas or mixtures of charged liquids), the virial pressure becomes:
\begin{align}
&P_\text{v}\left\{{\{\rho}_{i}\},\beta,\{{u}_{ij}\}\}\right\}
\nonumber\\
&=-\frac{1}{6}\int d^3r\left\{\sum_{i,j}{\rho_i \rho_j} h_{\text{eq},ij}(r)  r 
\frac{\partial u_{ij}(r)}{\partial r} \right\}\label{Pressure_virial_plasmas}
\end{align}
We substitute the solution to the DH Eqs.~\eqref{DH_2_k_11},\eqref{DH_2_k_22} and \eqref{DH_2_k_12} into Eq.~\eqref{Pressure_virial_plasmas}, written in the the Fourier space, and make use of a relation similar to Eq.~\eqref{Aux_1}. We repeat the steps analogous to Eqs.~\eqref{eq_charging_A_2_expl_2}, \eqref{eq_charging_A_2_expl_3} and \eqref{eq_charging_A_2_expl_3} and get:
\begin{align}
P_\text{v}\{{\{\rho}_{i}\},\beta,\{{u}_{ij}\}\} 
=&\int\frac{d^3k}{(2\pi)^3}\left\{
\sum_{i,j} \frac{\rho_i\rho_j}{2} h_{\text{eq},ij;k}\,\beta u^{\xi}_{ij;k}\right\}
\nonumber\\
&+\frac{4\pi}{3}\int^{\infty}_{0}\frac{dk}{\left(2\pi\right )^{3}}\left \{ \frac{k^{3}}{2\beta}\frac{\partial}{\partial k} \left[  \ln\left(D_{k}\right)
\right.\right.\nonumber\\&\left.\left.
-\beta \left (\rho_1 u_{11;k}+\rho_2 u_{22;k}\right )\right]
\vphantom{\frac{\partial}{\partial k}}
\right\}\label{Pressure_virial_2}
\end{align}
The first term in Eq.~\eqref{Pressure_virial_2}~can be identified from Eq.~\eqref{internal_energy_2} as the internal-energy density $U_\text{eq}\{\{\rho_{i}\},\beta,\{u_{ij}(r)\}\}/V$, calculated in the Fourier space. The second term in 
Eq.~\eqref{Pressure_virial_2} can  be integrated by parts (one may check that the integrand has correct behaviors at $ k=0$ and at $k \longrightarrow\infty$). We then get:
\begin{align}
&P_\text{v}\{{\{\rho}_{i}\},\beta,\{{u}_{ij}\}\} 
\nonumber\\
&=\frac{U_\text{eq}\{{\{\rho}_{i}\},\beta,\{{u}_{ij}(r)\}\}}{V}
\nonumber\\
&-\frac{1}{2\beta}\int\frac{d^{3}k}{\left(2\pi\right )^{3}}
\left\{\ln\left(D_{k}\right)-\beta \left (\rho_1 u_{11,k}+\rho_2 u_{22,k}\right )\right\}\label{Pressure_virial_3}
\end{align}
Finally, using Eq.~\eqref{eq_charging_A_2_expl_4} with $\xi=1$, we obtain:
\begin{align}
&P_\text{v}\{{\{\rho}_{i}\},\beta,\{{u}_{ij}\}\}
\nonumber\\ 
&=\frac{U_\text{eq}\{\{\rho_{i}\},\beta,\{u_{ij}(r)\}\}}{V}
-\frac{A_\text{eq}\{\{\rho_{i}\},\beta,\{u_{ij}(r)\}\}}{V}\label{Pressure_virial_4}
\end{align}

In order to calculate  the pressure from the thermodynamic definition, let us consider a volume in which there is one particle of the specie ``1'': $V_{1}=1/\rho_1$. Using the notion of this volume, we can calculate the thermodynamic pressure of the fluid as:
\begin{align} 
&P_\text{th}\{\{\rho_{i}\},\beta,\{u_{ij}\}\}
\nonumber\\
&=-\frac{d}{d V_{1}}\left[ 
\left(\frac{A_\text{eq}\{\{\rho_{i}\},\beta,\{u_{ij}(r)\}\}}{V}\right)V_{1}
\right] 
\nonumber\\
&=-V_1 \frac{d}{d V_{1}}\left[ 
\left(-\frac{A_\text{eq}\{\{\rho_{i}\},\beta,\{u_{ij}(r)\}\}}{V}\right)
\right]
\nonumber\\&\hphantom{=}-
\frac{A_\text{eq}\{\{\rho_{i}\},\beta,\{u_{ij}(r)\}\}}{V} 
\label{pressure_thermo_5}
\end{align}

However, in the volume $V_1=1/\rho_1$, there are also particles of the specie ``2''. Then, changing $\rho_1 $ by an infinitesimal value  $d\rho_1$, we also change $\rho_2$. Since the change of the density $\rho_1$ should not violate the overall neutrality of the plasma or charged liquids, we have to fulfill the relation $z_1\rho_1+z_2 \rho_2=0 $, where the charges $z_1, z_2$ are related to the asymptotic behavior of the interaction potentials at large distances $r$: $ u_{11}(r) \cong z^{2}_1/r $, $ u_{12}(r) \cong z_1 z_2/r $ and $ u_{22}(r) \cong z^{2}_2/r $. For this reason, the changes in $\rho_1$ and $\rho_2$
are not independent and we have:
\begin{equation} \label{derivative_rho}
\frac{d\rho_2}{d\rho_1}=\frac{\rho_1}{\rho_2}
\end{equation}
Then, the first term of the RHS of Eq.~\eqref{pressure_thermo_5} can be rewritten as: 
\begin{align}
-V_1 \frac{d}{d V_{1}}\left[ 
\left(-\frac{A_\text{eq}\{\{\rho_{i}\},\beta,\{u_{ij}(r)\}\}}{V}\right)
\right]
\nonumber\\
=\rho_1 \frac{d}{d \rho_{1}}\left[ 
\left(\frac{A_\text{eq}\{\{\rho_{i}\},\beta,\{u_{ij}(r)\}\}}{V}\right)
\right]
\end{align}
Denoting by $P^{(1)}_\text{th}\{\{\rho_i\},\beta,\{u_{ij}\}\}$ the first term of the last line in Eq.~\eqref{pressure_thermo_5} we thus have :
\begin{align}
P^{(1)}_\text{th}\{\{\rho_i\},\beta,\{u_{ij}\}\}
&=\rho_1 \frac{\partial}{\partial \rho_1}
\left(\frac{A_\text{eq}\{\{\rho_{i}\},\beta,\{u_{ij}(r)\}\}}{V}\right)
\nonumber\\ 
&+\rho_2 \frac{\partial}{\partial \rho_2}
\left(\frac{A_\text{eq}\{\{\rho_{i}\},\beta,\{u_{ij}(r)\}\}}{V}\right)
\label{pressure_thermo_6}  
\end{align}
However, as stems from Eq.~\eqref{eq_charging_A_2_expl_4},
$\beta A_\text{eq}/V$ depends on the $\{\rho_i\}$ only through the variables $\{\beta\rho_i\}$, i.e. we have $ \beta A_\text{eq}/V=\beta A_\text{eq}\{\{\beta\rho_{i}\},\{u_{ij}(r)\}\}/V$. For that reason we can write:
\begin{align}
&P^{(1)}_\text{th}\{\{\rho_i\},\beta,\{u_{ij}\}\}
\nonumber\\
&=\rho_1 \frac{\partial}{\partial \left (\beta \rho_{1}\right )}
\left(\frac{\beta A_\text{eq}}{V}\right)
+\rho_2 \frac{\partial}{\partial \left(\beta \rho_2\right)}
\left(\frac{\beta A_\text{eq}}{V}\right)
\nonumber\\
&=\frac{\partial\left(\beta \rho_{1}\right)}{\beta} \frac{\partial}{\partial \left(\beta \rho_{1}\right)}
\left(\frac{\beta A_\text{eq}}{V}\right)
+\frac{\partial\left(\beta \rho_{2}\right)}{\beta} \frac{\partial}{\partial \left(\beta \rho_2\right)}
\left(\frac{\beta A_\text{eq}}{V}\right)
\nonumber\\
&=\frac{\partial}{\partial \beta }\left(\frac{\beta A_\text{eq}}{V}\right)
=\frac{U_\text{eq}\{\{\rho_i\},\beta,\{u_{ij}\}\}}{V}
\label{pressure_thermo_7}
\end{align}
Eqs.~\eqref{pressure_thermo_5},\eqref{pressure_thermo_6} and \eqref{pressure_thermo_7} confirm the equivalence between the thermodynamic definition of pressure and the virial pressure formula in the case of the two-component DH theory based on the free-energy density functional proposed in the present paper.

\section{Conclusion}

Our Eqs.~\eqref{eq_A_form_postulated} and \eqref{eq_f_k} give an explicit expression of the  free-energy density functional in the Debye-H\"uckel (DH) approximation for two-component fluids, extending our previous results of \cite{Piron16}. These expressions allows one to obtain the DH equations from a minimization procedure with respect to the pair correlation functions. In the two-component DH case, as in the corresponding HNC case of \cite{Lado73b,Enciso87}, requiring that the exact charging relation is respected in the approximate model allows one to define a free-energy density functional that yields the correct expression for the internal-energy density and fulfills the virial theorem in the case of long-range potentials.

\acknowledgements
One of the authors (TB) acknowledges  funding from the EURATOM research and training program 2014--2018 in the framework of the ToIFE project of the EUROfusion Consortium under grant agreement No.~633053.

\bibliographystyle{unsrt}
\bibliography{biblio-utf8_TB_RP_100317}
\end{document}